# Software Engineering Modeling Applied to English Verb Classification (and Poetry)

Sabah Al-Fedaghi
Computer Engineering Department
Kuwait University
Kuwait
sabah.alfedaghi@ku.edu.kw

*Abstract*—**In requirements specification, software engineers create a *textual description* of the envisioned system as well as develop conceptual models using such tools as Universal Modeling Language (UML) and System Modeling Language (SysML). One such tool, called FM, has recently been developed as an extension of the INPUT-PROCESS-OUTPUT (IPO) model. IPO has been used extensively in many interdisciplinary applications and is described as one of the most fundamental and important of all descriptive tools. This paper is an attempt to understanding the PROCESS in IPO. The fundamental way to describe PROCESS is in verbs. This use of language has an important implication for systems modeling since verbs express the vast range of actions and movements of all things. It is clear that modeling needs to examine verbs. Accordingly, this paper involves a study of English verbs as a bridge to learn about processes, not as linguistic analysis but rather to reveal the semantics of processes, particularly the five "verbs" that form the basis of FM states: create, process, receive, release, and transfer. The paper focuses on verb classification, and specifically on how to model the action of verbs diagrammatically. From the linguistics point of view, according to some researchers, further exploration of the notion of verb classes is needed for real-world tasks such as machine translation, language generation, and document classification. Accordingly, this nonlinguistics study may benefit linguistics.**

## I. Introduction

This paper is concerned with fundamental notions such as *events*, *processes*, and *states* that have significance for progress in the field, especially in the areas of modeling in software engineering, artificial intelligence, and knowledge representation. Originally, *modeling* appeared as ontological schemes developed to understand the world based on fundamental entities and properties. However, contemporary physics discovered that all things have to be conceived fundamentally as PROCESS [1] (PROCESS is capitalized to distinguish it from the word *process* used in a different sense later in this paper). The fundamental way of describing PROCESS is with "activity verbs" [1].

Currently, one major scientific area that embraces modeling is software engineering. Software is everywhere in the infrastructure and affects all fields of life. Software engineers deal with more complex problems than any other engineering discipline [2]. Decades of work on software abstraction have helped gain intellectual control over systems of ever-increasing complexity. This mastery has motivated adopting a modeling approach throughout the software development process.

### A. Software engineering modeling

Requirements specification is a basic phase in software life cycle system development. Software engineers have put much effort into the process of transformation from requirements to software architecture, including creating a *textual description* of the envisioned system as well as *models*. The key problem is difficulty in giving an unambiguous, easy to understand description of a system and how it works. "We can do so with English descriptions; but such descriptions are often cumbersome, incomplete, ambiguous and can lead to misunderstandings" [3].

Specifically, *conceptual modeling* is performed by the requirements engineer to comprehend the problem domain and its requirements. Different models and various notations have been used, including Entity/Relationship Diagrams, Universal Modeling Language (UML), System Modeling Language (SysML), and Business Process Model and Notation (BPMN). For example, Rolland et al. [4] give a text specification and diagrammatic model of an ATM (Automated Teller Machine) as shown in Fig. 1. The model can be used in an early phase of system modeling to detect and understand problems and help clarify certain aspects of a system in more detail than just natural language.

One such tool, called the Flowthing Machine, FM, has been developed recently as an extension of the INPUT-PROCESS-OUTPUT (IPO) model. IPO has been used extensively in many interdisciplinary applications and is described as one of the most fundamental and important of all descriptive tools. As mentioned, the fundamental way of describing PROCESS is in verbs, and this has important implications for systems modeling since verbs express all the different actions and movements of all things. It is clear that modeling involves an examination of verbs.

- The user *inserts* the card.
- The system *checks* if the card is valid.
- A prompt for the code *is given*.
- The user *enters* the code.
- The system *checks* if the code is valid.
- A prompt "enter amount or select balance" *is given*.

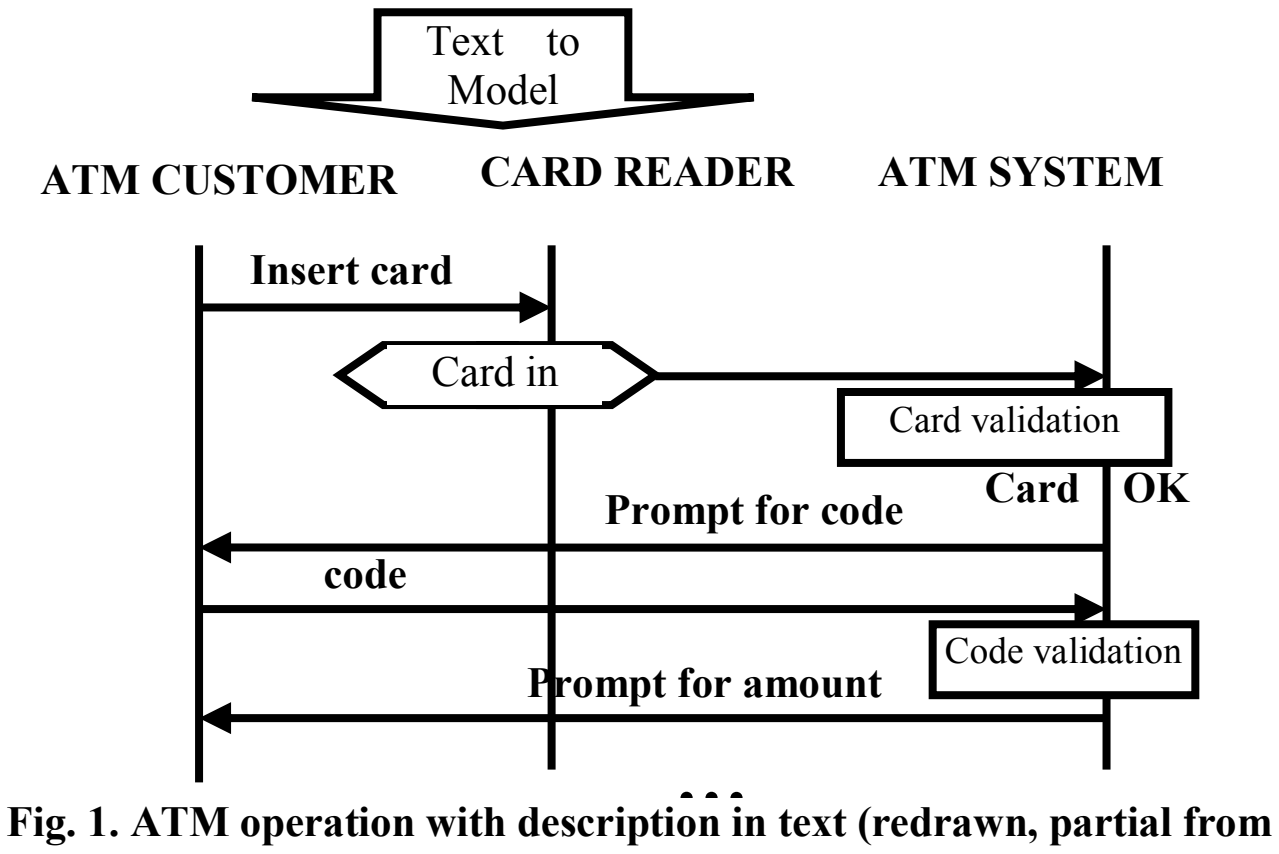

**Fig. 1. ATM operation with description in text (redrawn, partial from [4])**

Accordingly, this paper offers a study of English verbs as a bridge to learning about processes. The aim is not linguistic analysis; rather, it is to reveal the semantics of processes, particularly the five verbs used in FM modeling: create, process, receive, release, and transfer, and their relationship to verb classification, and specifically on how to model the action of verbs diagrammatically using create, process, receive, release, and transfer.

Using diagrammatic modeling to analyze use of English verbs in computer processes is not a new idea. Schalley [5] used UML to represent verbal semantics with diagrams. Fig. 2 shows a model of the action *wake up*. It reflects an extension of the UML to model the meaning of verbs, thus introduces "a third formal paradigm of computer science into linguistic semantics, one that is neither functional nor logical but object-oriented in nature" [5].

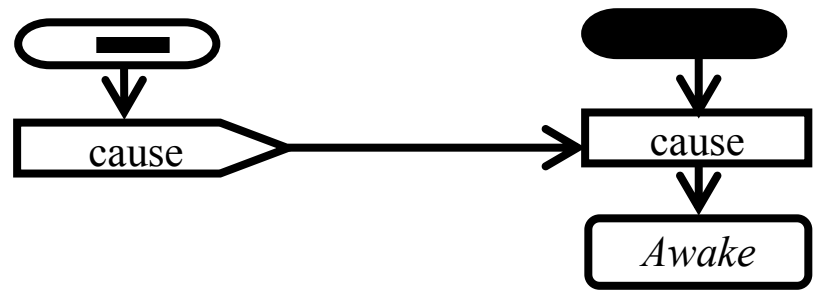

**Fig. 2. *Wake up* (redrawn, partial from [5])**

### *B. Verb classification*

Verbs often convey the main idea of a sentence [6]. They signify motion, and "every motion necessarily supposes some being or existence" [7]. They are "words which signify, to do, to be, or to suffer. They also express all the different actions and movements of all creatures and all things, whether alive or dead" ([7], attributed to [8]). In generative grammar, the verb plays a central role since it functions "as the nucleus in the deep structure, from which various surface utterances are processed" [7]. It is the most important element in the construction of utterances.

In the context of verbs, the focus in this paper is on *conceptualization* in terms of an abstract model of *things* that *exist* in a specific domain. Additionally, a diagrammatic language is adopted, since in computer science, "it is almost impossible to model without a conceptual diagram to visualize the modeler's concepts and the system" [9].

The concept of *verb* is closely related to *process*; in fact, process is sometimes viewed as a type of verb, or as a series of activities (i.e., verbs). According to Cousins [10], a *process* is "a set of interrelated or interacting activities which transforms inputs into outputs." Verbs are used to describe steps in a process (activities), and nouns are used to describe items output by activities to become input for other activities.

According to the English language site TESOL [11], *process* is a verb that indicates a change from one state to another. In general, verbs are classified into two types, dynamic and stative, as follows:

**Dynamic verbs**

- Activity: e.g., play, speak, run, and telephone. What we normally understand an "action" word to be.
- Process: e.g., ripen, change, and strengthen. To indicate a change from one state to another.
- Sensation: hurt, ache, and sting. Used to refer to bodily sensations.
- Momentary: e.g., knock, beat, and tap. Although closely related to the first category, these verbs have a shorter duration of action.

**Stative verbs**

- Cognition: e.g., know, remember, perceive, prefer, want, forget, and understand. These verbs have less to do with an overt action since they involve mental or cognitive processes.
- Perception: e.g., see, smell, feel, taste, hear. This small class of verbs is closely linked to verbs of cognition but centers on the senses rather than cerebral activity.
- Relational: e.g., be, consist of, own, have, seem, resemble, appear, sound, look (good), belong to. This category of verbs is used to connect two closely related concepts, usually through either equivalence or possession.

This gives a general idea of a textbook approach to verb classification. An enormous amount of work has been done in the field; to limit the problem to a manageable task, this paper focuses only on certain publications that lead to our goal: suggesting a diagrammatic tool that can be used in modeling verbs that could lead to a different approach to studying verbs.

### *C. Motivations*

According to Schuler [6],

> Despite the proliferation of approaches to lexicon (a place where all the information about the representation of words is stored [6]) development, the field of natural language processing has yet to develop a clear consensus on guidelines for computational verb lexicons, which has severely limited their utility in information processing applications… Resources such as verb lexicons are frequently language and domain specific, not always available to the whole community, and are expensive and time-consuming to build.

Verb classes can be used in tasks such as machine translation, language generation and document classification. A further exploitation of the notion of verb classes is needed for real-world tasks [6]. Accordingly, the diagrammatic tool introduced in this paper may benefit research in this area.

A more direct motivation is that this paper is an attempt to more fully explain a recently introduced conceptual model (FM model) that has been used in several areas [12–16]. The FM model comprises five "verbs" acting in a flow machine: *transfer*, *process* (*existing* things), *release*, *receive*, and *create things*, as shown in Fig. 3. The claim in the FM representation is that these five verbs are basic operations in any system, physical or otherwise. The claim in such a model is that *all "verbs" can be mapped (or reduced) to create, process, release, transfer and receive!* Such a claim needs more extensive investigation, hence the study of verb classification might shed light on this point. We will model different examples of verbs and their behaviors that have puzzled researchers in English and translate these examples into create, process, release, transfer, and receive and see the results with the aim of understanding limitations of the FM proposition.

To provide background on the FM model, a brief description is given in section 2. The sections that follow apply these diagrams to examples from the literature in linguistics.

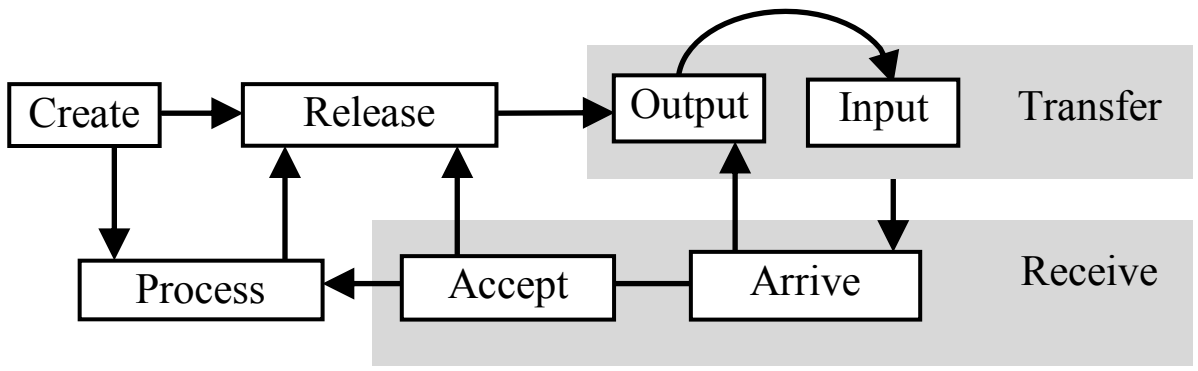

**Fig. 3. Flow machine**

## II. FLOWTHING MACHINE

This section briefly reviews FM, which forms the foundation of the theoretical development in this paper; however, the examples given here are new contributions.

### A. Basic notions

The FM model (see [12–16]) is a diagrammatic schema that depicts the existence of *flow things* (hereafter, *things*), defined as *what can be created*, *released*, *transferred*, *received*, and *processed*, by means of stages in a flow machine (Fig. 3). *Things* begin to flow through the stages of the machine when they are created by the machine or imported from other machines.

*Flow* here entails transition or realization of change and movement and positioning. **Create** is the emergence of a thing in the system from outside it. The rest of the flow is relinquishment of one stage for the next one. Such flows are specified in an analogy of drawing traffic flows on a city map. Then, as will be discussed later, *dynamic* flows are added in terms of *events* that describe the behavior of the system. In this latter case, the streets of the city become streams of flow of cars, people, etc.

The point here is that a *flow* is often thought of as physical movement, but in FM, it can be much more than that. It is a notion that also captures *conceptual* movement in thought, sensation, being, and doing. The modeler builds a conceptual construct and also conceptual "movement," which we call flow. Thus, a physical house *flows* from a sphere (e.g., *class* in UML terminology) to another class when there is a transition from a person owner to a certain bank, and a car has various *flows* to robots and workers *simultaneously* when it is processed, e.g., one fixes glass while another changes tires. Flows might be fast or slow, parallel or sequential, physical or digital (e.g., uploading software) or mental (e.g., inspecting finished products), only creating, only processing, etc.

The stages in Fig. 3 can be described as follows:

**Arrive**: A thing reaches a new machine.

**Accept**: A thing is approved to enter a machine. If arriving things are always accepted, *Arrive* and *Accept* can be combined as a **Receive** stage.

**Process** (change – close to TESOL's [11] *process as a type of verb* discussed in the introduction): The *thing* goes through some kind of transformation that changes its "state" without creating a new thing.

**Release**: A thing is marked as ready to be transferred outside the machine. Note that things can be released from a given system without being transferred, as in the case of sent emails waiting for a damaged channel to be fixed.

**Transfer**: The thing is transported somewhere from/to outside the machine.

**Create**: A new thing is born (created) in a machine.

Flow machines use the notions of *spheres and subspheres*. These are constructs (mental products) of machines and submachines. Multiple machines can exist in a sphere if needed. A sphere can be a person, an organ, an entity (e.g., a company, a customer), a location (a laboratory, a waiting room), a communication medium (a channel, a wire). A machine is a subsphere that embodies the flow; it itself has no subspheres. This notion of a sphere is taken from cognitive linguistics where an *idea* is treated as a complex unit that is associated with other entities or other forms of association. "A door, for example, also connotes a door knob, a key hole, a door jamb, etc." [17].

FM also utilizes the notion of *triggering*. Triggering is the activation of a flow, denoted in machine diagrams by a *dashed arrow*. It is a dependency relationship among flows and parts of flows. A flow is said to be triggered if it is created or activated by another flow (e.g., a flow of electricity triggers a flow of heat), or activated by another point in the flow. Triggering can also be used to initiate events such as starting up a machine (e.g., by remote signal). Multiple machines can interact by triggering events related to other machines in those machines' spheres and stages.

### A. Examples

From a linguistic point of view, create, process, release, transfer, and receive can be thought of (logically) as *predicates* while the *thing* is the subject. There are many ways of classifying English verbs, e.g., verbs of movement, verbs of appearance, verbs of disappearance, verbs of Existence, etc. [18]. Samples of these types are modeled as follows:

**Verbs of Putting**: Consider an example of what Levin [18] calls the class of "Verbs of Putting," e.g., *I put the book on the table*. Fig. 4 shows the FM representation of this statement. The verb is designated as a sequence of discrete operations that form a subset of {create, process, receive, release, transfer}. It can be interpreted as: I release and transfer (output) the book (the thing) to be transferred and received (input) on the table. Accordingly, *put* has been "dissolved" in release, transfer and receive—OR the sequence of predicates *Release*(book) *Transfer*(book) *Transfer*(book) *Receive*(book). According to Levin [18], a verb such as as *put* refers to "putting" an entity at some *location*.

Fig. 4 is a static representation of *I put the book on the table.* It is what we previously referred to as a "city map." Behavior is modeled by considering events over this static description. This notion will be illustrated in the next example.

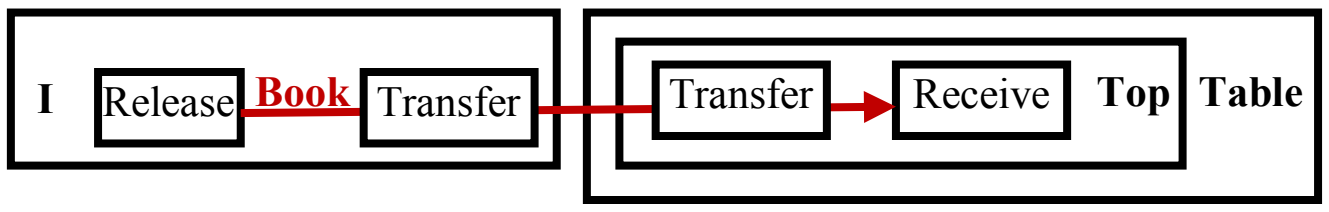

**Fig. 4. FM representation of *I put the book on the table***

**Verbs of Removing**: *Doug removed the smudges from the tabletop* [18]. See Fig. 5. We notice that the past tense *removed* indicates an event completed in the past. An event is a *thing* that can be created, processed, received, released, and transferred in *time*. Time is a *thing* that can also be created, processed, received, released, and transferred. An event has its "space" comprising the components of time and itself. Accordingly, Fig. 6 shows the event *Doug removed the smudges from the tabletop*. Note that when time is released and transferred, this indicates a past event. Also, *Process* of an event (top flow in Fig. 6) indicates that an event *runs its course*.

**Verbs of Sending and Carrying**: *Nora sent the book to Peter*. See Fig. 7.

**Verbs of Exerting Force:** *Nora pushed the chair*. See Fig. 8.

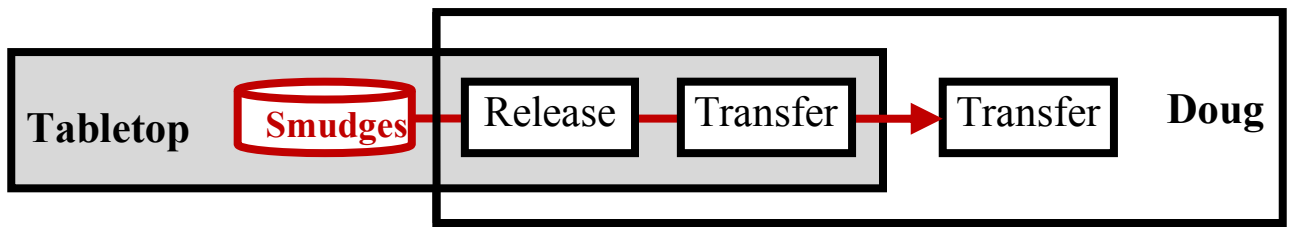

**Fig. 5. Representation of *Doug removed the smudges from the* table top**

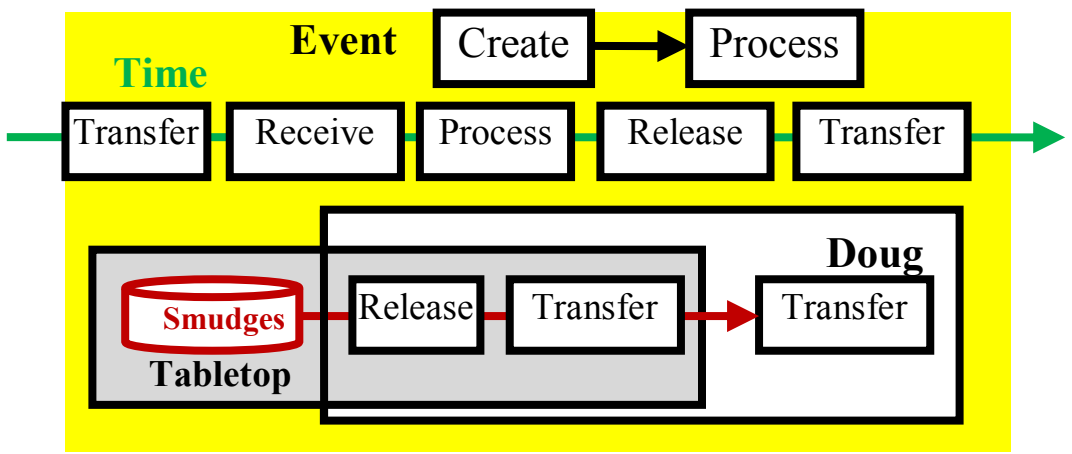

**Fig. 6. The event *Doug removed the smudges from the table top.***

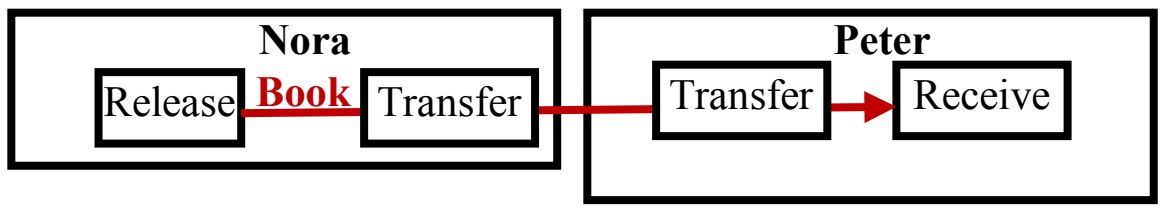

**Fig. 7. FM representation of *Nora sent the book to Peter.***

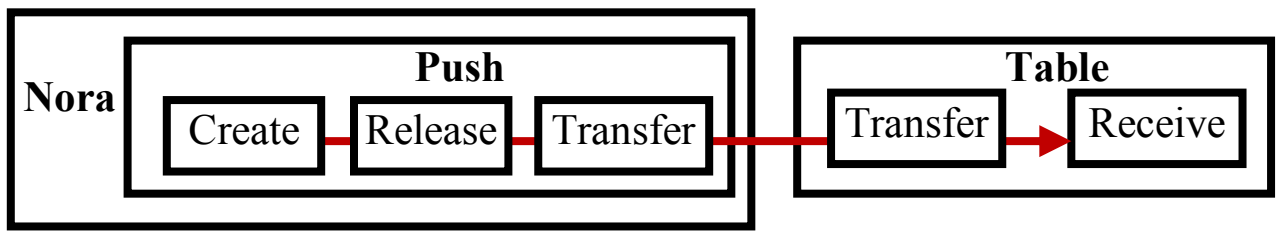

**Fig. 8. FM representation of *Nora pushed the chair.***

**Verbs of Change of Possession**: *They lent a bicycle to me*. See Fig. 9. Since an example of an event has been given, we don't show the event version of this statement.

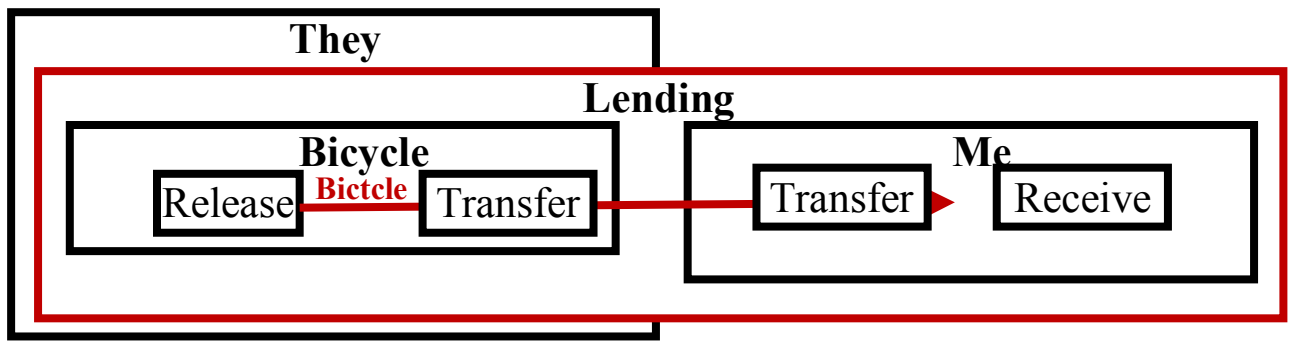

**Fig. 9. FM representation of *They lent a bicycle to me.***

**Verbs of Learning:** *Rhoda learned French from an old book.* See Fig. 10. *Process* in the context of flow means *change*. Thus, the verb is depicted as a sequence of discrete operations: transfer (output), transfer (input), receive, and process (change in knowledge).

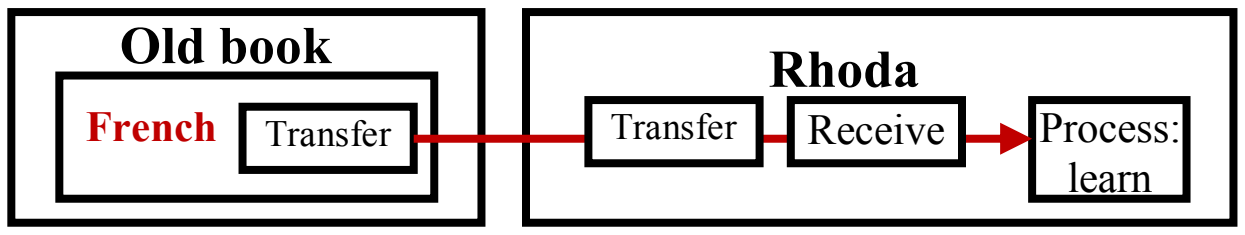

**Fig. 10. Representation of *Rhoda learned French from an old book.***

**Verbs of Holding and Keeping**: *She held the rail*. See Fig. 11.

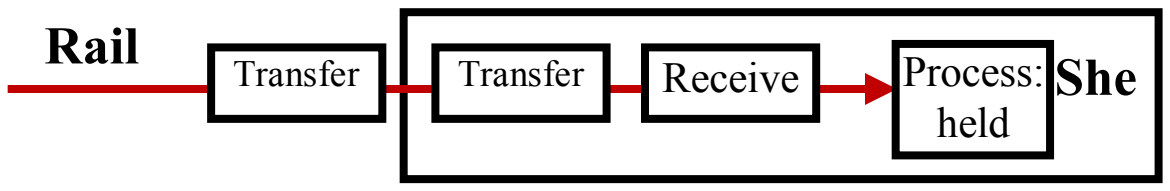

**Fig. 11. Representation of *She held the rail.***

**Verbs of Concealment:** *Frances hid the presents from Sally*. See Fig. 12.

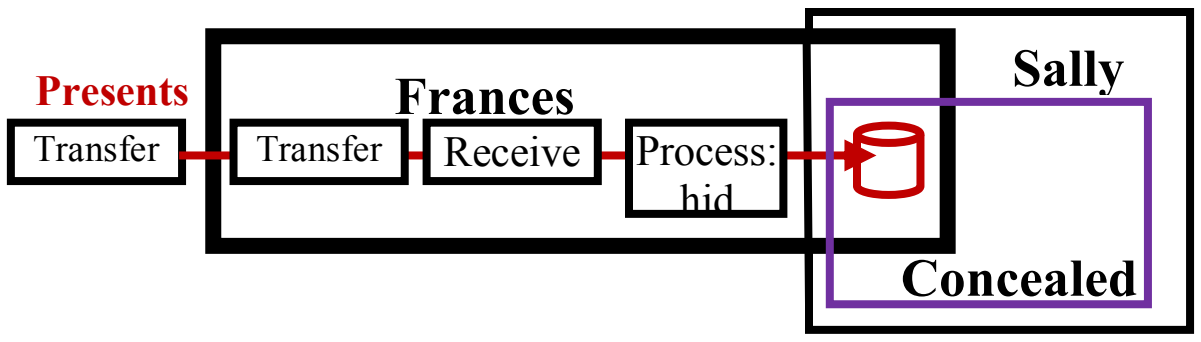

**Fig. 12. Representation of *Frances hid the presents from Sally.***

These examples demonstrate the expressive strength of the FM language and its capability to present different types of verbs. The claim here is that such a diagrammatic language plays the same role that it provides for software engineers: to give an unambiguous, easy to understand description of a scheme and how it works. Again, "We can do so with English descriptions; but such descriptions are often cumbersome, incomplete, ambiguous and can lead to misunderstandings" [3].

Consider Levin's [18] analysis of advice verbs:

(1) *Ellen warned Helen.*

(2) * *Ellen warned to Helen*. (The "*" indicates incorrect English [18])

(3) *Ellen warned (Helen) against skating on thin ice*.

"These verbs relate to giving advice or warnings. The verbs in this class are among the verbs in English that allow a PRO-arb (see [18]) object interpretation when used intransitively. The exception is the verb *alert*, which requires an obligatory object" [18]. Fig. 13 shows the three expressions in diagrammatic form. These diagrams provide another way to look at the expression and thus could enhance the analysis.

In *Ellen warned Helen*, Helen is processed (being warned) by Ellen. As mentioned previously, flow in FM does not indicate a physical flow; rather, it means that Helen as a conceptual thing comes under the sphere of Ellen to be warned. In **Ellen warned to Helen*, the "*to*" indicates "sending" a thing (warning) to Helen. (Again, the "*" indicates incorrect English [18])

### B. *Example: poetry*

This subsection applies the FM model in a larger context than that of one-statement diagrams. The aim is to further demonstrate the expressive power of the model.

Henry Wadsworth Longfellow (1807–1882) was a Harvard scholar, poet, and novelist. His lyric poem *The Arrow and the Song* compares *shooting an arrow* and *singing a song*; both are lost in the air but are found again, the arrow in an oak tree and the song in the heart of a friend.

I shot an arrow into the air,
It fell to earth, I knew not where;
For, so swiftly it flew, the sight
Could not follow it in its flight.

I breathed a song into the air,
It fell to earth, I knew not where;
For who has sight so keen and strong,
That it can follow the flight of song?

Long, long afterward, in an oak
I found the arrow, still unbroke;
And the song, from beginning to end,
I found again in the heart of a friend.

The arrow, a weapon, could represent our destructive behavior. A song suggests something carefree and benign. According to Nield [19] in analyzing the poem, "We can never predict the power of our actions. The word said, the deed done, disappear into the past, but often, years later, we can be astounded to learn of their impact. A friend explodes with rage over an imagined slight; a stranger thanks us for a favor we'd forgotten."

Fig. 14 shows the *static* FM representation of the poem. It is a construct in the modeler's mind made up of things, spheres, and flows regardless of their nature, e.g., physical, mental, or even fantasy. It includes four principal spheres: I (circle 1 in the figure), Air (2) Earth (3), and those with sight so keen and strong (4). I shot an arrow (5): Retrieved (transferred/received) an arrow and processed (shot) it. The arrow flew (6), generating (creating) (7) a flight (8) in the air. Note that, for simplicity, the arrow machine of the *arrow itself* is not surrounded by a box since it is recognized from the flow.

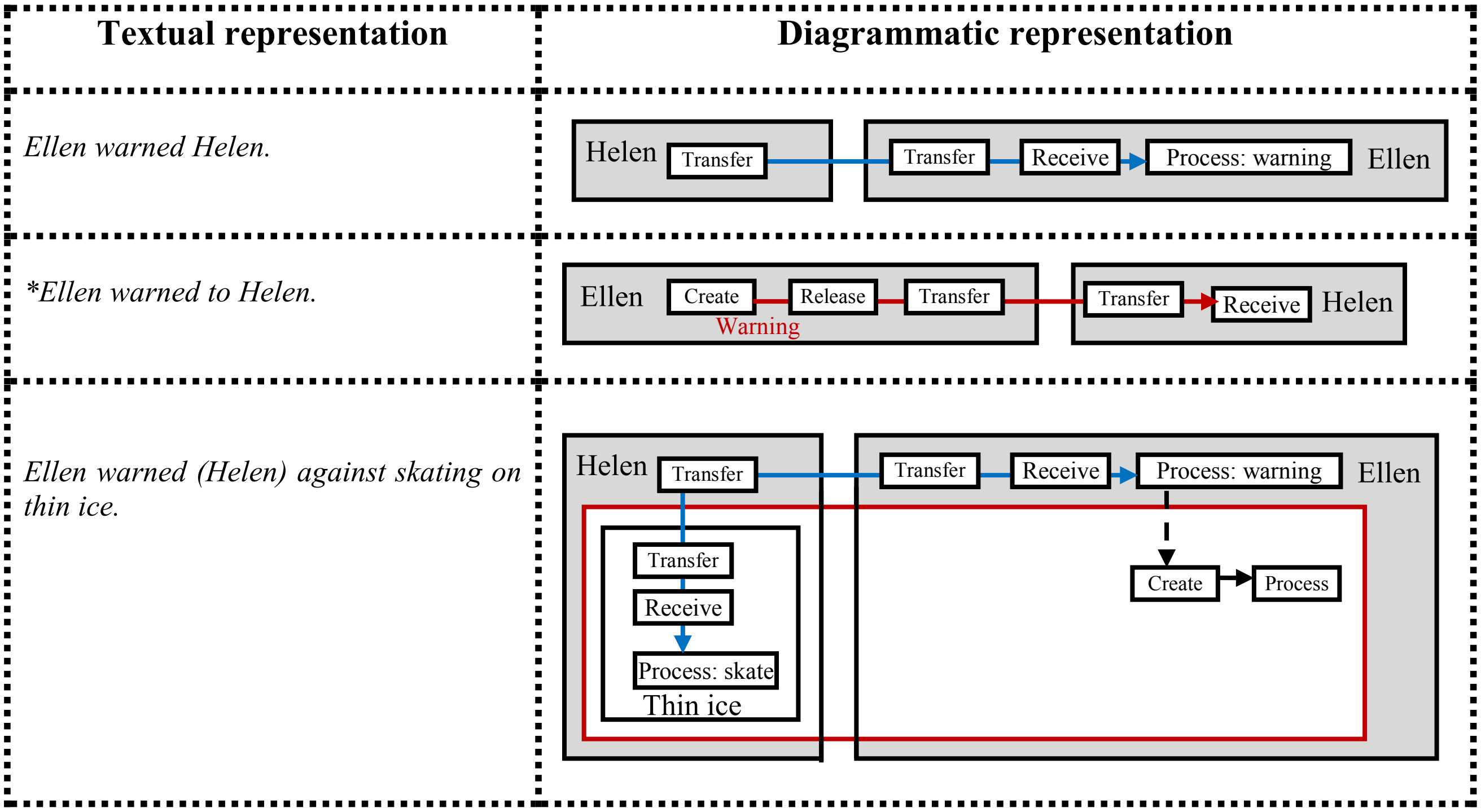

**Fig. 13. FM representations of the three expressions of an advice verb.**

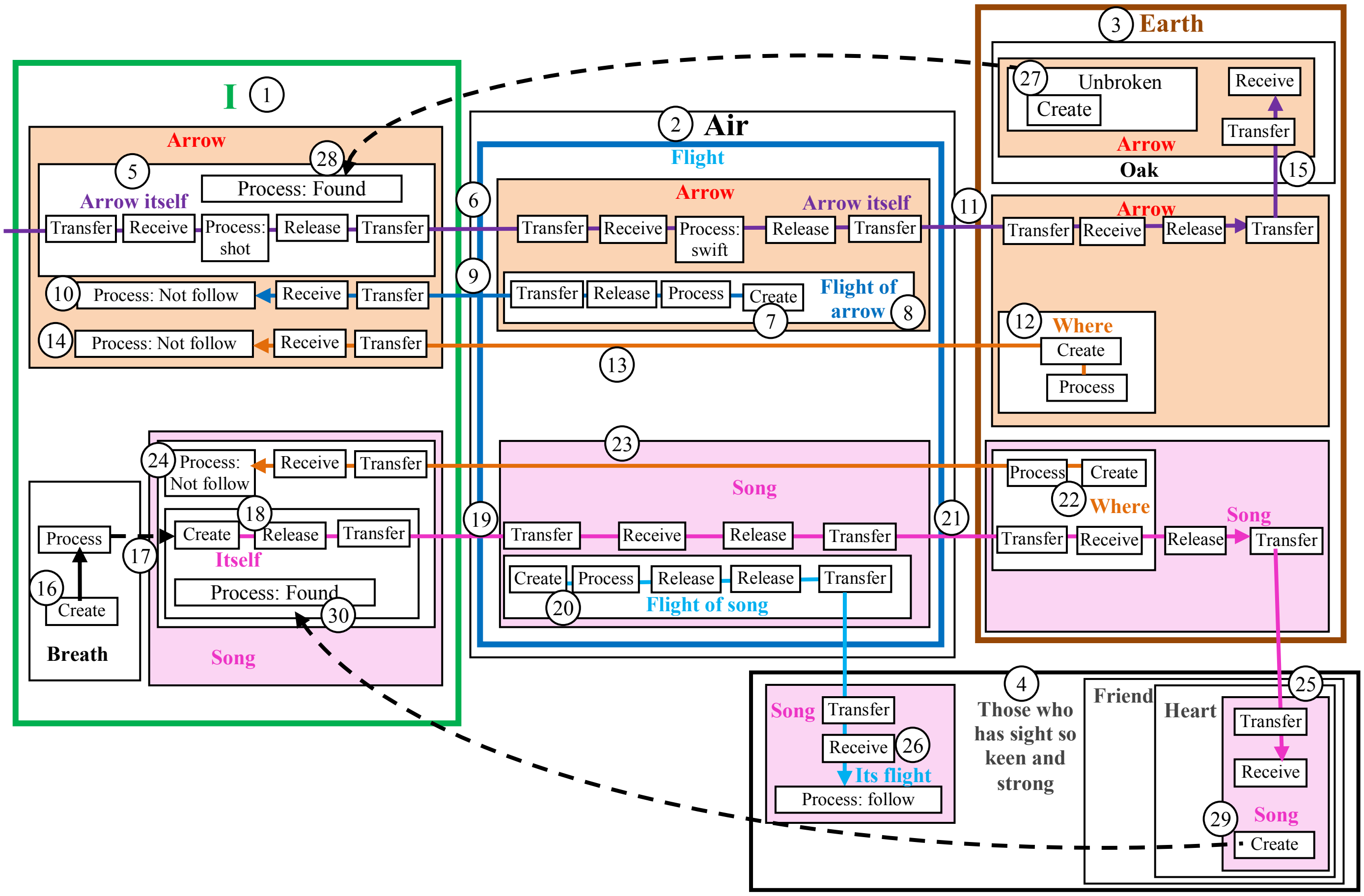

**Fig. 14. The FM representation of the poem**

The flight itself is cognized by me, and this awareness of the flight comes (flows) to me (9), but I cannot process it to the level of following it (10). The arrow flows to the earth (11) in some location (where, 12); however, in spite of the flow of this location awareness cognized by me (13), I cannot process it to the level of knowledge of exact location (14); however, as I will find later, the arrow has landed in an oak (top right corner, 15).

My breath (lower left corner, 16) triggers (17) a song (18) that flows (19), generating (creating) (20) a flight in the air. The song flows to the earth (21) in some location (where, 22); however, in spite of this the flow of location awareness cognized by me (23), I cannot process it to the level of knowledge of its exact position (24); however, as I will find later, the song has landed in a heart (lower right corner, 25). The *flight itself* of the song arrives (flows) to those with sight so keen and strong (26).

Long, long afterward (this is modeled at the dynamic level of the model of the poem), the arrow *appears* (is created, 27) in the oak, as I come to find out (upper curved dashed arrow, 28). And the song *appears* (is created, 29) in the heart of those who have sight so keen and strong (bottom curved dashed arrow, 30).

The diagram exposes different aspects of the poem at different levels. Consider for example the contrast between the arrow that (in general) represents destructive behavior and the song that points to gentle action. The diagram highlights that the arrow (most likely) *is not made* by the speaker, while the song is *created* by him/her. An immature critic (e.g., the author of this paper) might suggest mirroring the two actions; i.e., the poet "should" somehow have emphasized that he is a "maker" of the arrow. The point here is that the diagrammatic representation exposes the anatomy of the poem, thus opening the door to all types of comprehensions and remarks.

However, this is not the purpose of the diagram; rather, it aims to demonstrate the expressive power of the FM representation. The dynamism of the poem can be modeled by execution of the sequence of operations embedded in the poem (events). Let us select the following eleven events:

**Event 1**: I shot an arrow into the air
**Event 2**: It flew too swiftly to see
**Event 3**: It fell to earth
**Event 4**: I knew not where

**Event 5**: I found the arrow in an oak
**Event 6**: I breathed
**Event 7**: I created and transferred a song [into the air]
**Event 8**: Who has sight so keen and strong, can follow the flight of song?
**Event 9**: It fell to earth
**Event 10**: I knew not where
**Event 11**: I found the song in the heart of a friend

Fig. 15 shows the map of some of these events laid over the static description of Fig. 14. Fig. 16 shows the chronology of all events, and Fig. 17 shows the first five events. Note that it is possible at this point to distinguish between the operational sequence and the temporal sequence.

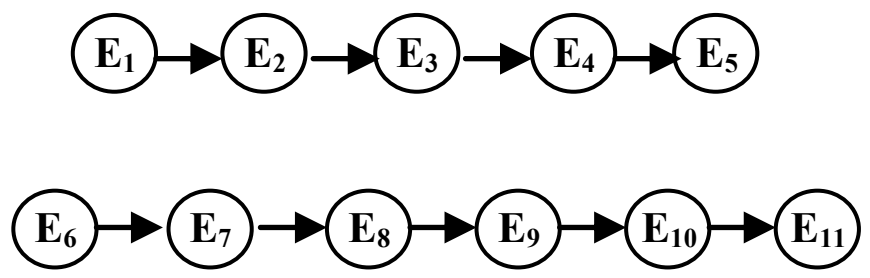

**Fig. 16. Chronology of events**

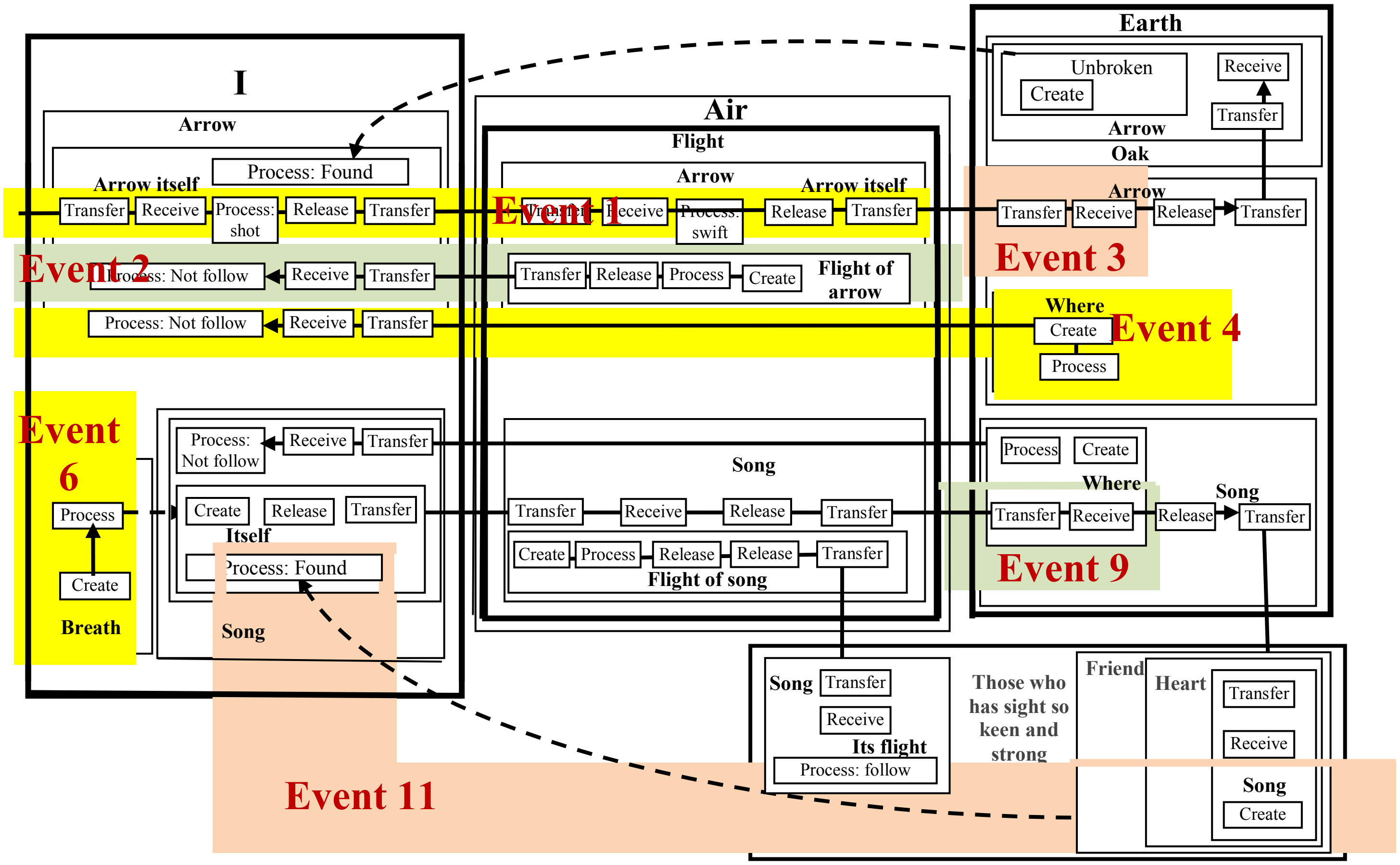

**Figure 15. The FM representation of the poem**

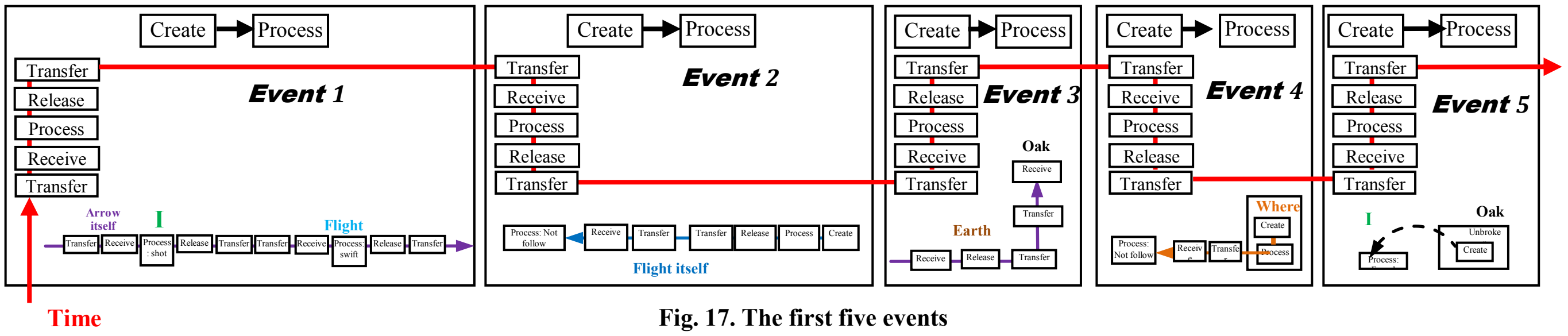

**Fig. 17. The first five events**

## III. APPLYING FM TO *TENSE*

Verbs involve the concept of time by differentiating among past, present, and future. In English there are three grammatical patterns that relate to time and reality: Tense, Aspect, and Mode. *Tense* refers to the way verbs change their form to reflect at which time an event takes place. It is "the grammatical means through which speakers conceptualize and encode time" [20]. This section applies FM to tenses.

Note that the aim here is twofold:

- To show that all verbs can be mapped to *Create*, *Process*, *Receive*, *Release,* and *Transfer* or a sub-set of them, and
- To show that all tense *forms* are describable in terms of these five verbs in the spheres of *Time*, *Event,* and *static description*.

### A. Present, past, and future

***I walk*** (Fig. 18a): In the figure *I generate (create) walking*. Here *Create* denotes the appearance of the walking phenomenon in the world. Note that there is no indication that the walking has been completed (no Release and Transfer of time) or that the walking is going on (no Process of walking). Processing Time indicates *Now* (Time of event). Accordingly, the diagram can be interpreted as:

*Now, there is a process that has created a situation or phenomenon of walking.*

***I walked*** (Fig. 18b): In the figure showing an event-ized description, *Now* is time after the completed event of walking,.

***I will walk*** (Fig. 18c): In the figure the *Now* (time process) precedes the walking event, whose time has not yet arrived.

### B. Progressive tense

According to Lecercle [21], "The '-ing' suffix is one of the glories of the English language. Because it has a double origin a mark of the present participle, and the mark of a series of nouns." For example,

***I am walking***: In Fig. 19, the "whole" event of a walk (circle 1) is still going on since it has not yet entered the *Release* and *Transfer* stages of the Time sphere (2). However, walk as a "unit" of walking (e.g., a step) is being repeated (3) as a sub-event (4). Each sub-event (e.g., a single step) has its time and finishes (5). This can be seen as analogous to a film being shown that has not finished. Each frame (picture) of the film comes and goes as a sub-event of the ongoing film event.

Similarly, **I was walking** *and* **I will be walking** *involve adding sub-events* (not shown in figures).

### C. Perfect tenses

According to Dowty [22], "aside from the progressive [tense], no English tense has received more attention from linguists and yet eluded a convincing analysis so completely as the present perfect." The perfect present represented by *I have finished washing the dishes* is described as follows:

- We normally use the Present Perfect when we want to talk about something which happened in the past but is relevant now.
- We use the Present Perfect to show a direct link with the present.

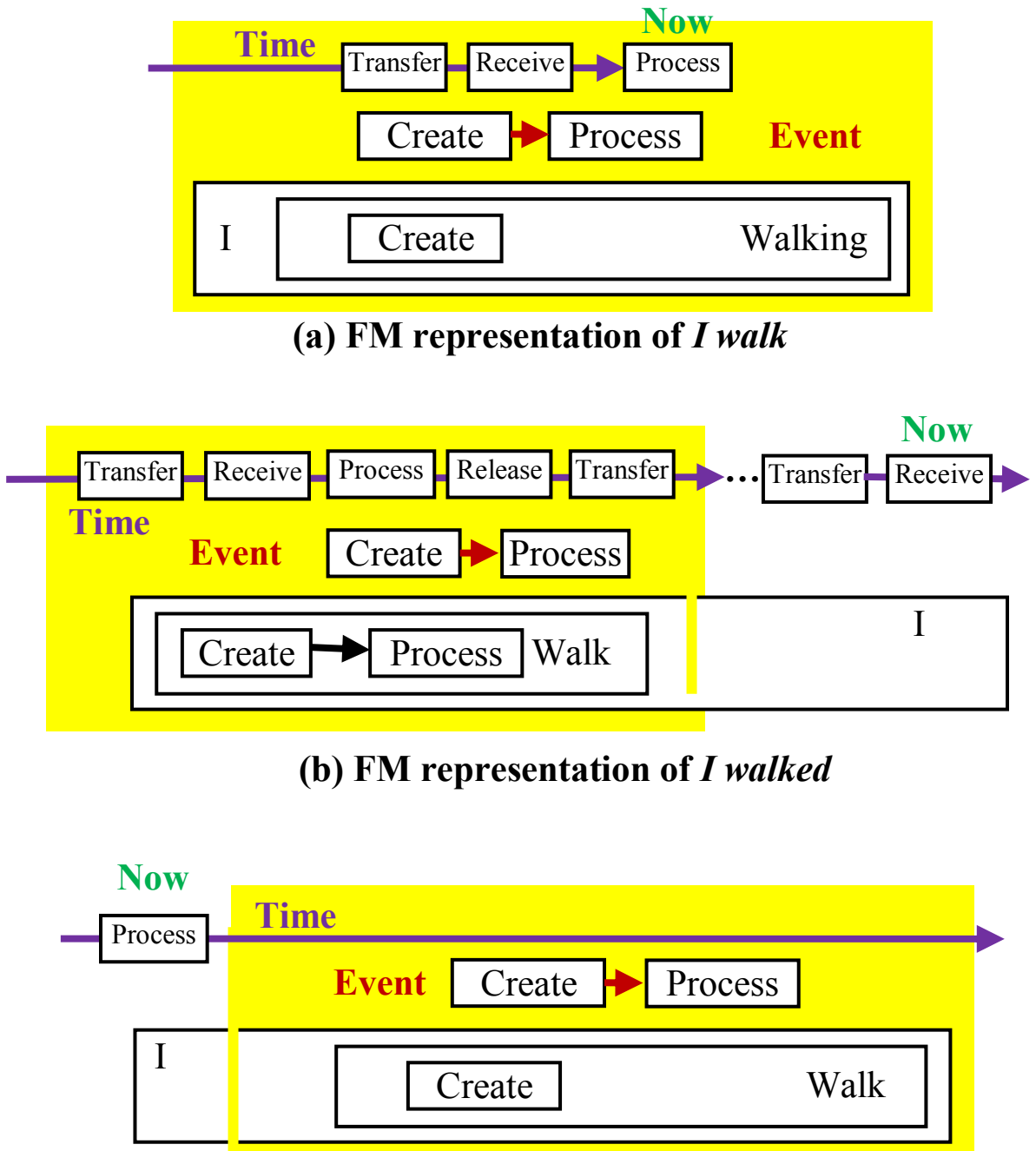

**(a) FM representation of *I walk***

**(b) FM representation of *I walked***

**(c) FM representation of *I will walk***

**Fig. 18. FM representations of present, past, and future**

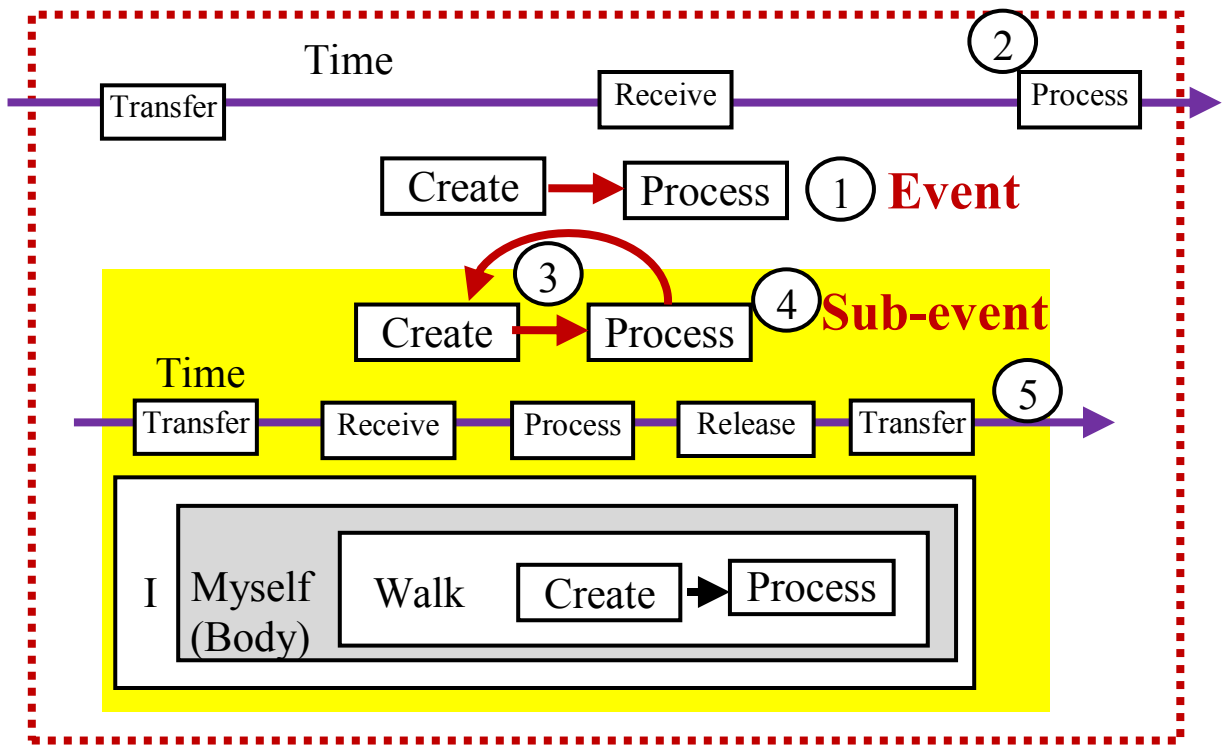

**Fig. 19. FM representation of *I am walking***

- We use it for something that happened in the past BUT when the present result is important. e.g: I think I have eaten something bad. I don't feel well. [23]

According to Foohs [24], in the present perfect location of events the time reference is indefinite. Simple past tense, on the other hand, narrows down the temporal location of a prior event to some well-defined limit.

> The defining function of the perfect in English is to express the pastness of the event embodied in the lexical verb, together with a certain applicability, pertinence, or relevance of the said past event(s) to the context of the speech, the "now" of the speaker or writer. The simple past, in contrast, appears when the event in hand is past but lacks the connection of relevance to the present. [24]

Without a claim to any contribution to linguistics, we speculate that our diagrammatic representation might help shed light on this issue of the perfect in English. Fig. 20a shows the FM representation of *I have washed the dishes*. First the sphere of *I* (circle 1) includes the sub-sphere *Have* (2) containing the event (3) of the dishes being washed (4). The event occurred in the past and I claim that I have this event in the present time (as I speak *Now*). Here, according to the author's interpretation, the neutral auxiliary "have" is taken in its *literal* sense of "to own" to mean that, e.g., *I have won the race* is a claim of ownership of an event. According to this interpretation *I have washed the dishes* could be seen as a declaration of ownership ("have") of an event..

The same interpretation can be applied to past perfect, *I had washed the dishes*, but the "having" of the event was in the past, as shown in Fig. 20b. Fig. 20c shows the representation of the future perfect *I will have washed the dishes*.

Regardless of the acceptance of these unconventional "interpretations", the point is that these sentences can be represented in terms of the five FM stages.

## IV. Activity and Accomplishment

Yet a verb can also indicate other ways in which that verb involves the notion of time. According to Vendler [25],

> Verbs have tenses indicates that considerations involving the concept of time are relevant to their use… Distinctions have been made among verbs suggesting processes, states, dispositions, occurrences, tasks, achievements, and so on… These differences cannot be explained in terms of time alone… Nevertheless one feels that the time element remains crucial.

Vendler [25] introduced four categories of verb classification:

(1) *Activity* terms (activities denote ongoing *dynamic* situations): run, walk, swim, push (a cart), drive (a car), etc.
(2) *Accomplishment* terms (accomplishments and achievements both express a *change* of state): paint (a picture), make (a chair), build (a house), run (a mile), walk (to school), deliver (a sermon), etc.
(3) *Achievement* terms: reach (the summit), win (the race), die, find, ...
(4) *State* terms (i.e., *static* situations): have, desire, love, hate, want, know, believe, rule, etc. [26]

In spite of the fact that "it has been frequently pointed out that his classification has some difficulties" [26], the taxonomy still has "a significant influence" on linguistic research and philosophical literature, "with many refinements that extend types of verbs into more than four categories" [26].

> Vendler breathed new life into an old Aristotelian tripartition of situational types by proposing a quadripartition: States, Activities, Accomplishments, and Achievements… More specifically, Vendler's [25] proposal seems to incorporate the claim that the category of verbs of any natural language can be split up into these four categories. [27]

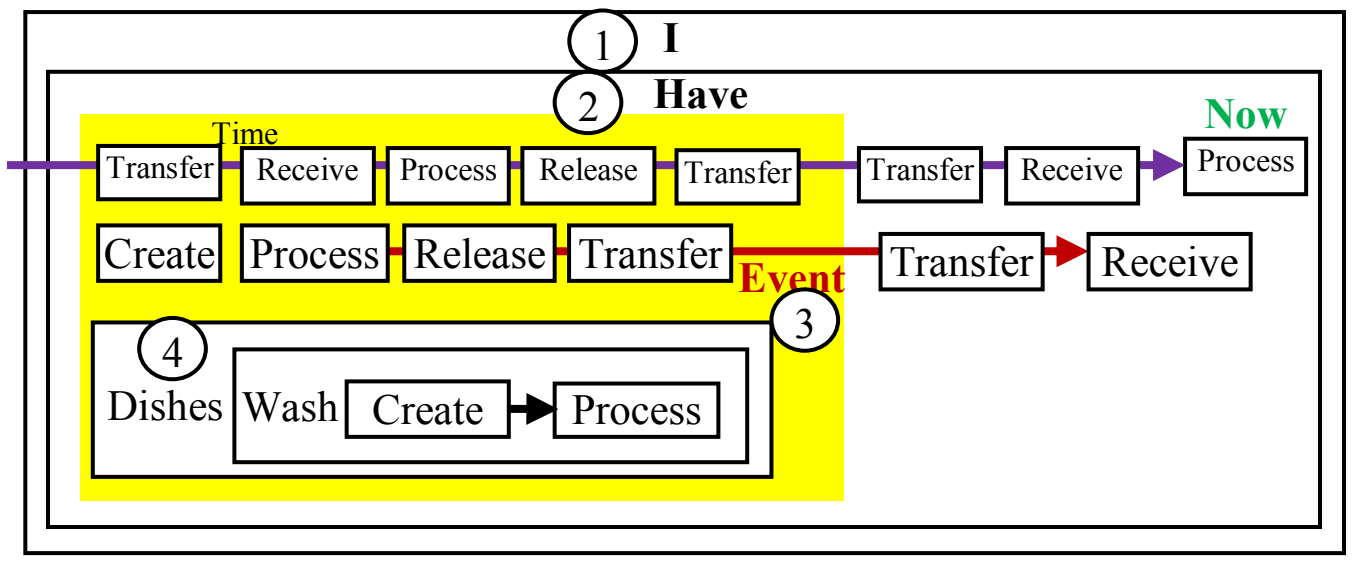

**(a) FM representation of *I have washed the dishes***

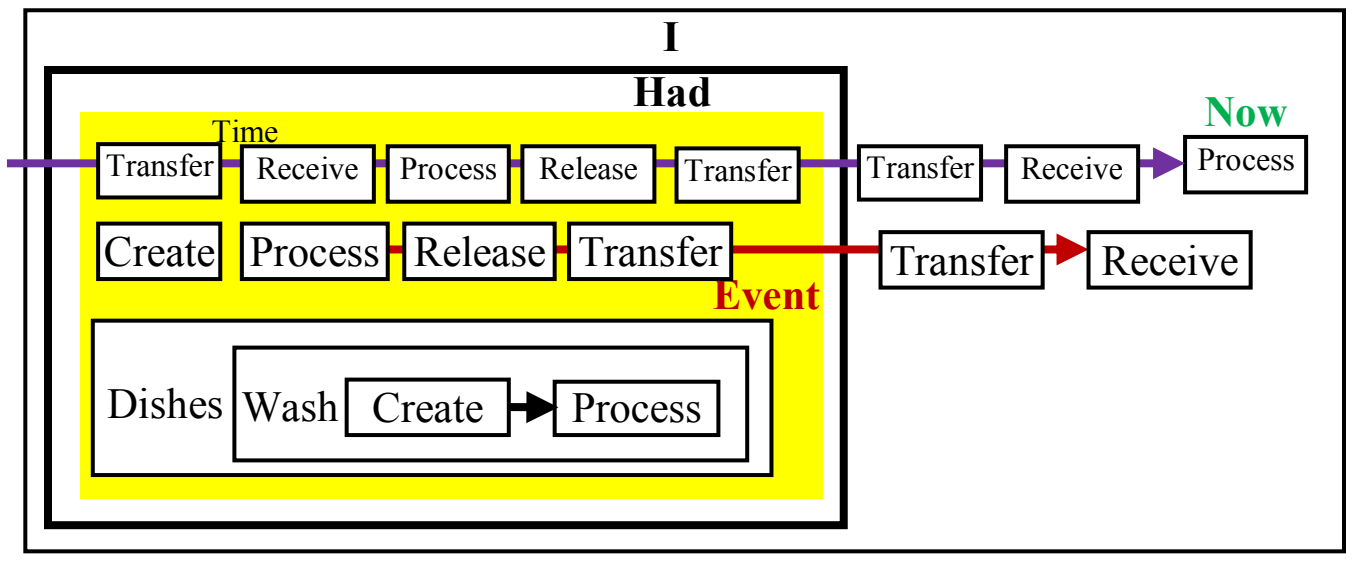

**(b) FM representation of *I had washed the dishes***

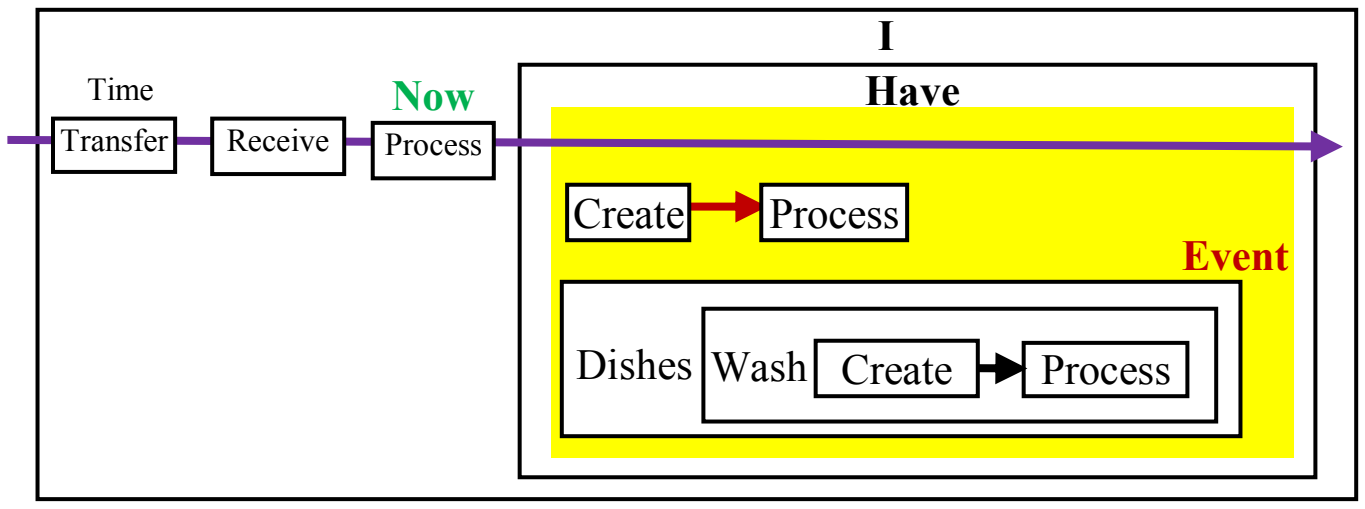

**(c) FM representation of *I will have washed the dishes***

**Figure 20. Perfect tense**

This section presents some examples of FM models of activity and accomplishment. According to Vendler [25], there is a difference between *running* and *running for a mile*. *Running, pushing a cart,* and so forth are *activity* terms whereas *running a mile*, *drawing a circle*, and so forth are *accomplishment* terms.

### *A. Activity vs. accomplishment: Example 1*

**Activity***:* *I am running* [26]: If I say that someone is running or pushing a cart, my statement does not imply any assumption as to how long that running will go on; he might stop the next moment or he might keep running for half an hour [25].

Fig. 21a shows the static (with respect to time) FM representation of *I am running*. It is a statement or declaration with no implications about time. In the figure, *I* myself (body) *create* running (as an action) and *process* it (engage in it). This type of action is typically described as "*running* is *about* the runner." In FM this indicates that the runner *processes* his/her physical self. Note in such a view that *Running* is a *thing* that can be created, processed, etc. Thus the verb *am running* describes a flow of participants: the things *Running* and *I*. Other participants (or things) *Time* and *Event* may enter to convert the static description to behavior, as will be explained next.

As in the case of progressive tenses previously discussed with the statement *I am walking* (Fig. 19), the event-ized version of *I am running* is shown in Fig. 21b. The event (circle 1) has its time (2) that has not finished since it is in the *Process* stage of time (3) and has not yet flown to the *Release* and *Transfer* stages. Paraphrasing Vendler [25], no assumption is implied as to how long that running will go on.

Fig. 21b also shows that this event includes a sub-event that is repeated in time (4–5). In this picture, "pieces" of running (sub-events) are performed repeatedly (i.e., instances of moving forward) in a continuous manner and reoccur as each sub-event is created and processed (runs its course) in the effort to reach a complete running (the whole event). The picture here is more complicated than Vendler's [25] description: it does not imply any assumption as to how long the "running unit action" will be repeated. Each "running unit action" is completed, but the repeating process is not complete.

**Accomplishment**: Fig. 22 shows the static representation of *He is running a mile*, where "one will keep running till he has covered the mile … running a mile does have a 'climax,' which has to be reached if the action is to be what it is claimed to be" [25]. Note that a mile is a space that "receives" the person and "releases" him/her.

Note that *He is running a mile* does not imply *change* (e.g., in the runner's position). The change occurs when it happens as an *event*. Events are changes in things; *He* changes his position and *the run* is created and processed.

The time factor is introduced in Fig. 23. As previously in *I am running* (Fig. 21b), here again there is the repeated sub-event (1) of "running unit" and the whole running event (2); however, the sub-event is repeated within the mile (3). Finishing the mile (4) triggers finishing the total running (5).

In comparing the activity diagram *I am running* and the accomplishment diagram *He is running a mile*, it is clear that there is a difference, a difference best left to linguists to explain. Our aim has been accomplished: to provide a diagrammatic tool for understanding and explaining the problem involved.

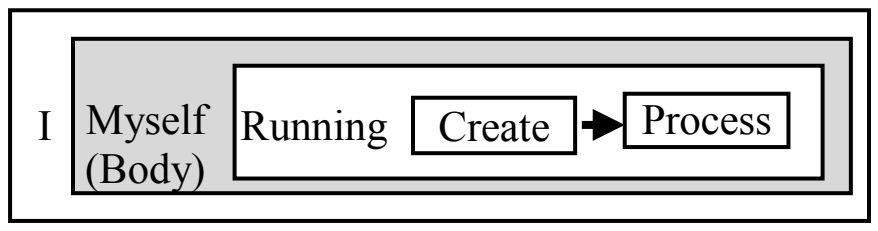

**(a) Static representation of *I am running***

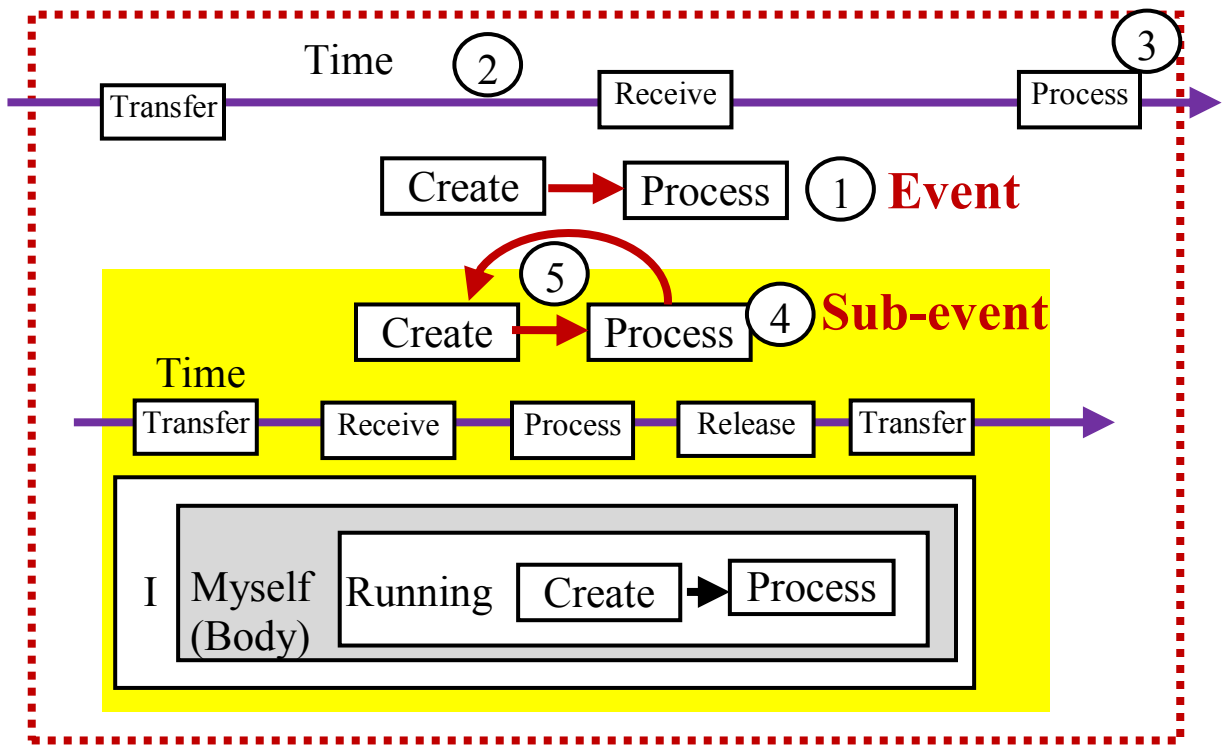

**(b) The event *I am running***

**Fig. 21. Representations of *I am running***

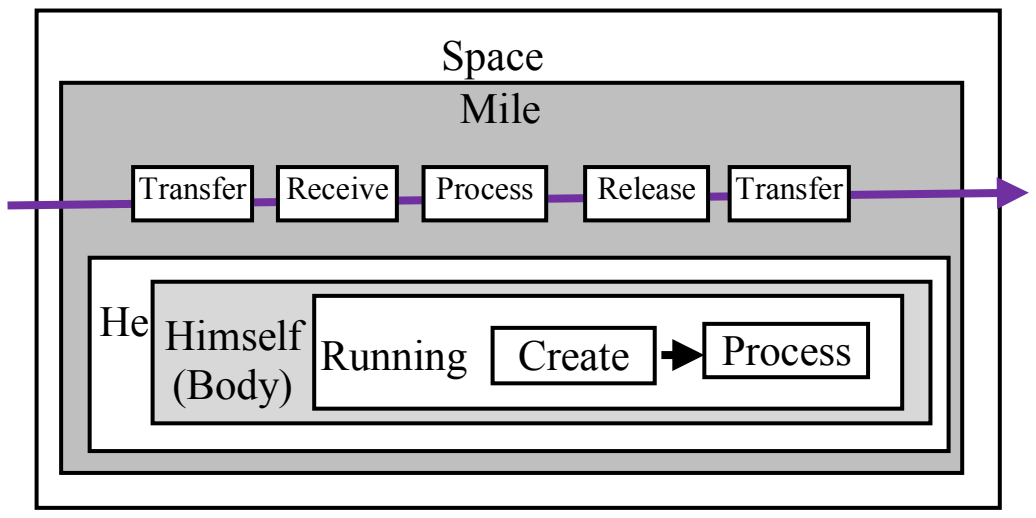

**Fig. 22. FM representation of *He is running a mile***

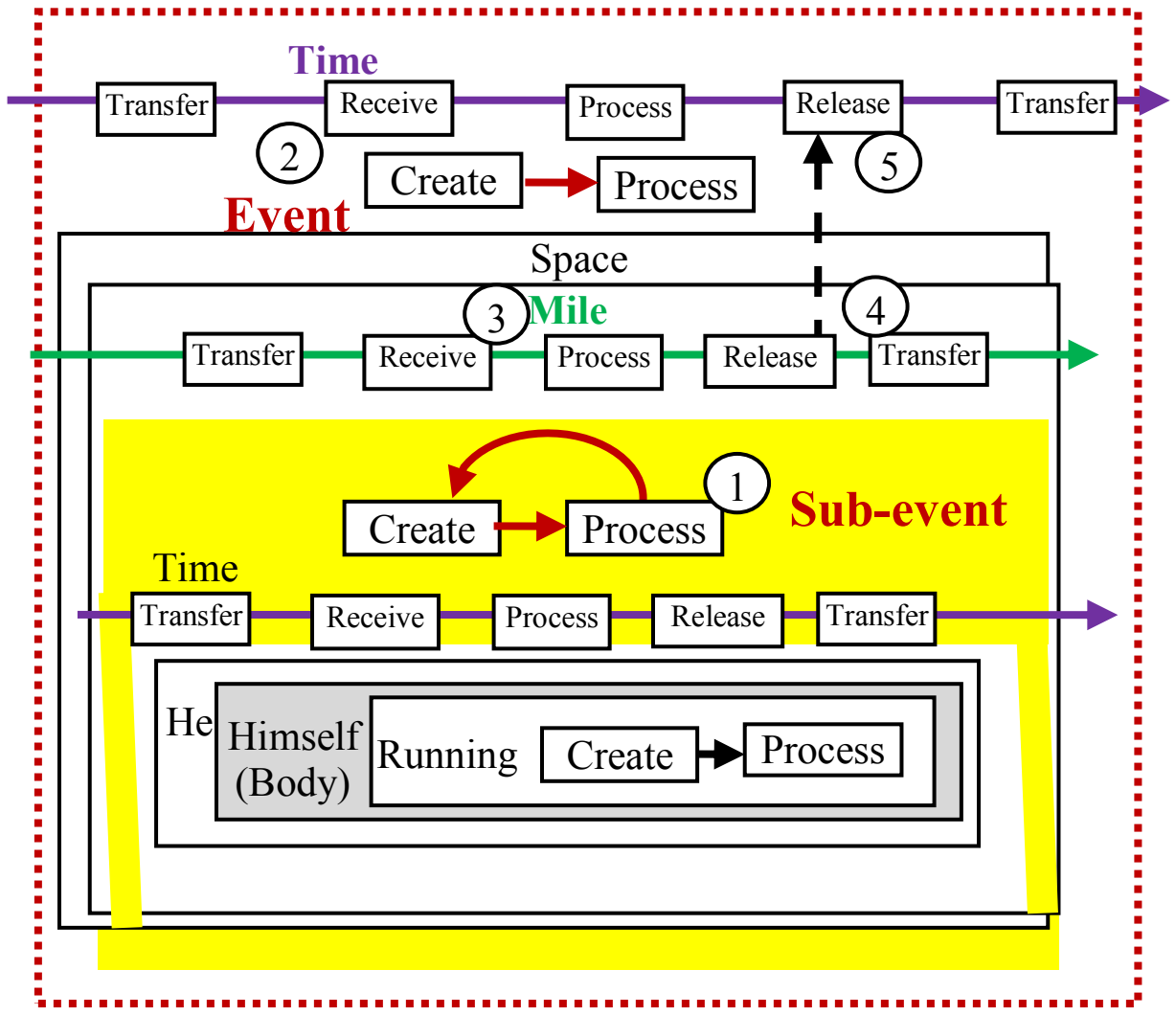

**Fig. 23. Event representation of *He is running a mile***

*B. Additional example of accomplishment*

***I am writing a letter*** [26]: To make this example more current, we modify it to *I am writing an email*. Fig. 24a shows the FM representation of this case. *I*, in the sub-sphere of my *Email*, create the email (a piece of writing) and process it..

Fig. 24b shows *I am writing an email* as an event. It includes two events:

- Event 1 is creation of an email (1). This event is complete since its time machine has stages of Release and Transfer that follow Process, the accumulated moments of creation that occur in Event 2:
- Event 2 is a repeated sub-event that follows Event 1 and comprises the *processing* of this email (3). This sub-event is performed repeatedly (6) as pieces of writing are created and processed. This generation of writing is a continuing sub-event as Event 2 in the stage of Process of time (7) since it has no Release or Transfer of time.

The question here is, what is a common characteristic between *running a mile* and *writing an email*?

*C. Activity vs. accomplishment: Example 2*

According to [26], activities and accomplishments are distinguished by the kind of adverbials they are compatible with. As stated by Kawamura [26], interpreting Vendler [25], "accomplishments do and activities do not have a set terminal point which is logically necessary to their being what they are [26]."

**Activity:** *He pushed the cart for half an hour* [26]. Fig. 25a shows a model of the static description in which a person receives a cart and processes it, which involves pushing it. Fig. 25b shows an event that lasts half an hour. It includes two sub-events:

- Receiving a cart (2)
- Pushing the cart (3), performed repeatedly (4).

Note that the times of the two sub-events are not included because this information is immaterial for the analysis.

**Accomplishment**: *He drew the circle in twenty seconds* [26]. Fig. 26a shows its static description and Fig. 26b shows the model of its event. *He drew the circle in twenty seconds* involves an event that (1) has two sub-events (2 and 3). Sub-event 1 involves repeatedly drawing (4) until a circle is created (5).

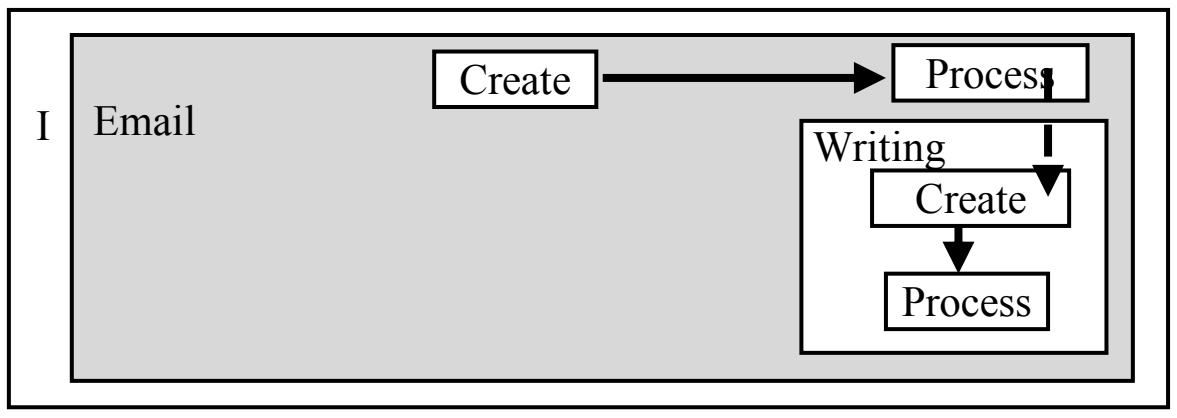

(a) FM representation of *I am writing an email*

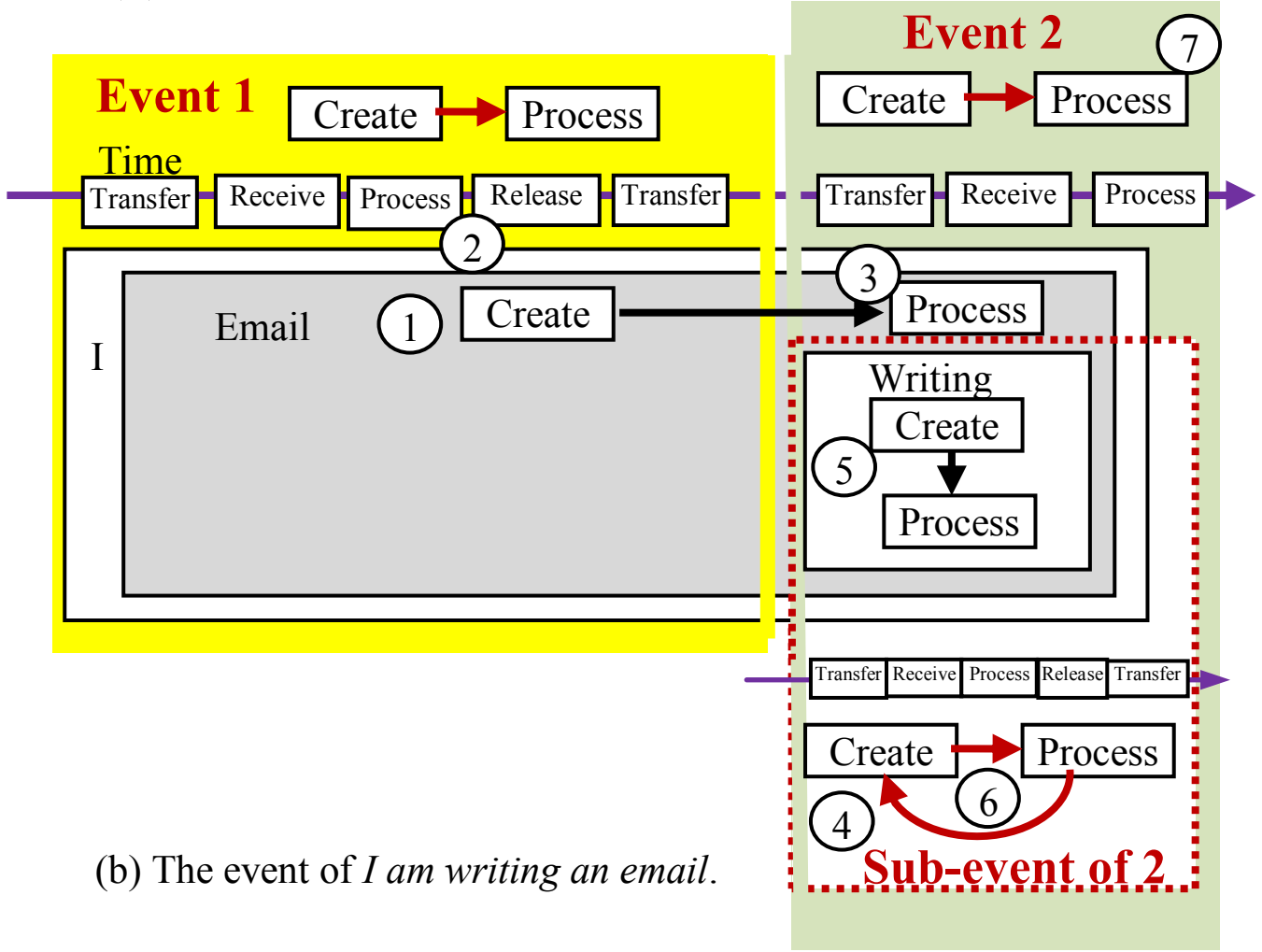

(b) The event of *I am writing an email.*

**Fig. 24. FM representations of *I am writing an email* and its event**

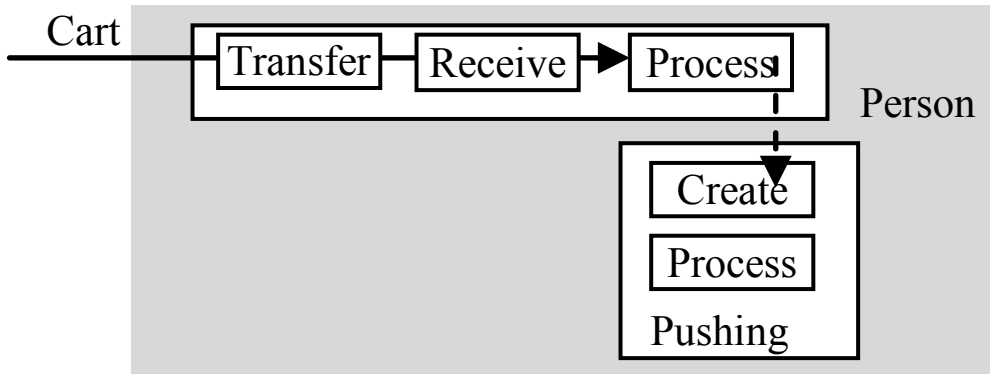

(a) Static representation of *He pushed the cart for half an hour*

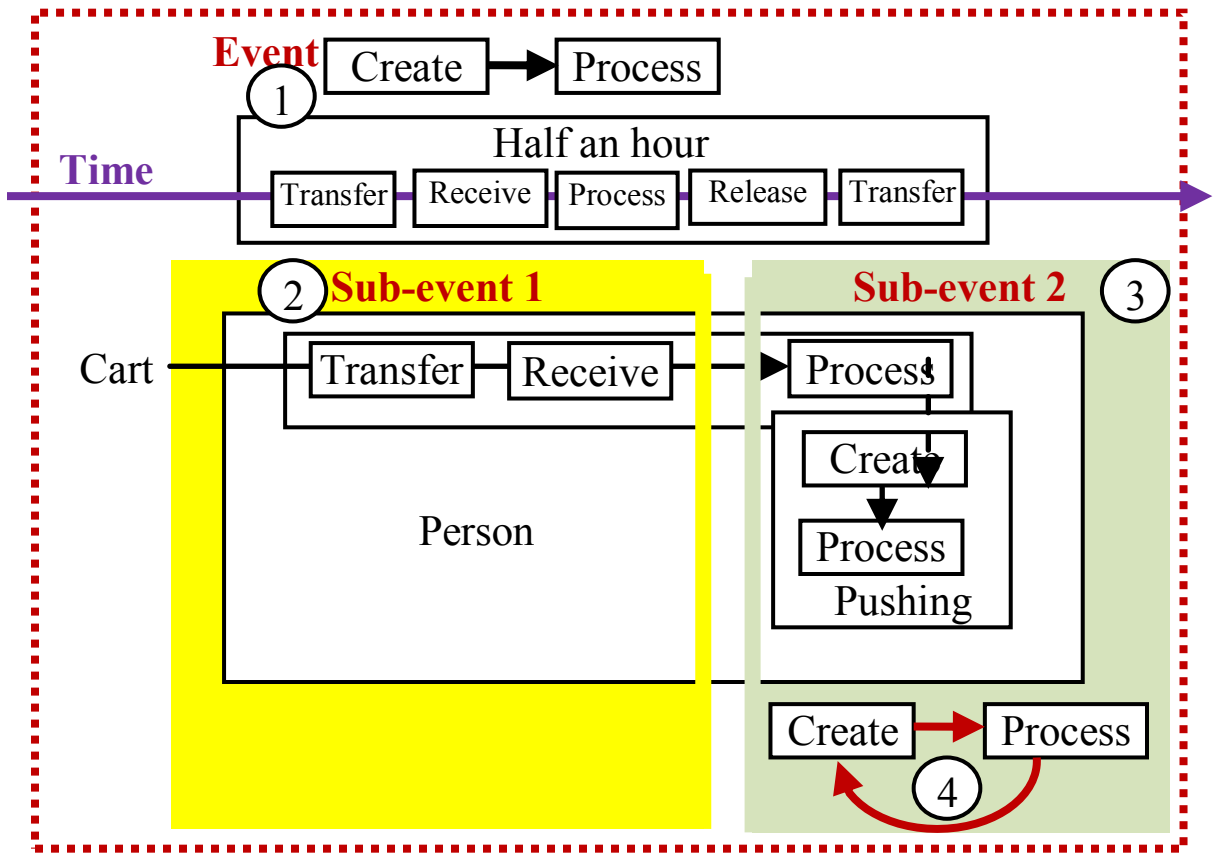

(b) The event *He pushed the cart for half an hour*

**Fig. 25. *He pushed the cart for half an hour* and its event**

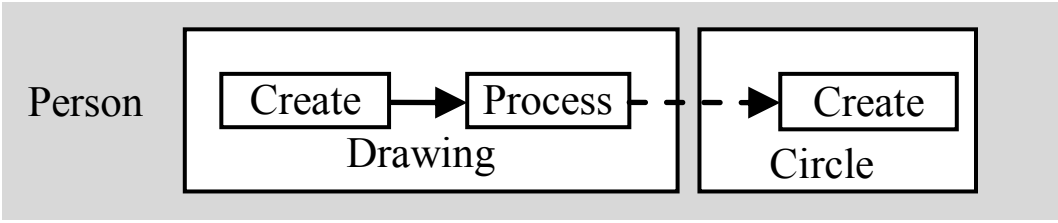

**(a) The event of *He drew the circle in twenty seconds***

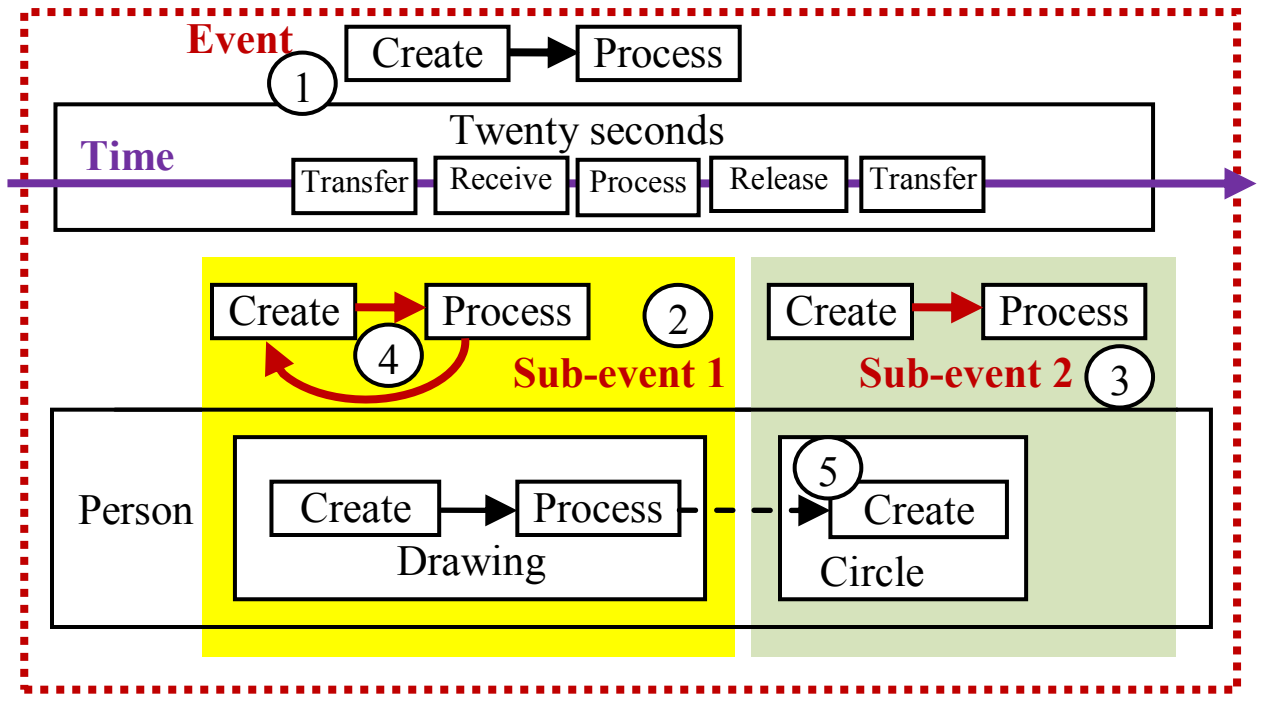

**(b) The event *He drew the circle in twenty seconds***

**Figure 26. *He drew the circle in twenty seconds* and its event.**

## V. Conclusion

This paper has explored the five verbs of the FM model: Create, Process, Receive, Release, and Transfer. FM is an extension of the input-process-output model that has been used in many interdisciplinary applications. Hence, understanding the "verb connection" to its extension FM seems to have important implications for systems modeling. As a by-product of that, it is proposed to use diagrammatic modeling as a tool to analyze English verbs and as another way to look at the verbal expressions that may enhance such an analysis. The results demonstrate that FM can express English verbs diagrammatically.